\documentclass[12pt]{article}
\usepackage[centertags]{amsmath}
\usepackage{amsfonts}
\usepackage{amssymb}
\usepackage{amsthm}
\usepackage{newlfont}
\usepackage{color}
\newtheorem{lemma"}{Lemma}[section]

\def \n {\noindent}

\usepackage{fancyhdr}
\begin{document}

\begin{center}
{\bf{\color{red}  January 2015}}\\
=====\\

*****\\

 {\color{blue}{\Large \bf On an application of generalized Jentzsch theorem to Gribov operator in Bargmann space}}\\

\end{center}

\begin{center}
*****
\end{center}
\begin{center}
 {\it Abdelkader Intissar}
\end{center}

\begin{center}
{\it - Equipe d'Analyse spectrale , Facult\'e des Sciences et Techniques, Universit\'e
de Cort\'e, 20250 Cort\'e, France}\\
{\it T\'el: 00 33 (0) 4 95 45 00 33}\\
 {\it Fax: 00 33 (0) 4 95 45 00 33}\\
 {\it intissar@univ-corse.fr}\\
 \end{center}
 \begin{center}
{\&}
\end{center}
{\it - Le Prador, 129, rue du Commandant Rolland, 13008 Marseille,
France}
\begin{center}
*****
\end{center}

{\color{blue}{\Large\abstract}}{\it In Bargmann representation, the reggeon's field theory{\color{blue} [5]} is caracterized by the non symmetrical Gribov operator\\

$\displaystyle{H_{\lambda',\mu,\lambda} = \lambda' A^{*^{2}}A^{2} + \mu A^{*}A + i\lambda A^{*}(A + A^{*})A}$\\

where $A^{*}$ and $A$ are the creation and annihilation operators;  $[A, A^{*}] = I $.\\
$(\lambda',\mu, \lambda) \in \mathbb{R}^{3}$ are respectively the four coupling, the intercept and the triple coupling of Pomeron  and  $i^{2} = -1$.\\

For $\lambda' > 0 ,\mu > 0$, let $\sigma (\lambda',\mu) \neq 0$ be the smallest eigenvalue of $H_{\lambda',\mu,\lambda}$, we show in this paper that $\sigma (\lambda',\mu)$ is positive, increasing and analytic function on the whole real line with respect to $\mu$ and that the spectral radius of $H_{\lambda',\mu,\lambda}^{-1}$ converges to that of $H_{0,\mu,\lambda}^{-1}$ as $\lambda'$ goes to zero.\\ The above results can be derived from the method used  in ({\color{blue} [2]} Commun. Math. Phys. 93, (1984), p:123-139) by Ando-Zerner to study the smallest eigenvalue $\sigma (0,\mu)$ of $H_{0,\mu,\lambda}$, however as $H_{\lambda',\mu,\lambda}$ is regular perturbation of $H_{0,\mu,\lambda}$ then its study is much more easily. We can exploit the structure of $H_{\lambda',\mu,\lambda}^{-1}$ to deduce the results of Ando-Zerner established on the function $\sigma (0,\mu)$ as $\lambda'$ goes to zero.\\}

\n{\bf Acknowledgments}: The author would like to express his thanks to Professor Shinichi Mochizuki for its suggestions and comments on the french version of this paper.\\

$\overline{\quad \quad \quad \quad \quad \quad \quad \quad \quad \quad \quad \quad \quad \quad \quad \quad \quad \quad \quad \quad \quad \quad \quad \quad \quad \quad \quad \quad \quad \quad \quad \quad \quad \quad \quad \quad \quad \quad \quad \quad \quad \quad}$ \\
\n {\bf Keywords:} Spectral analysis; Non self-adjoint Gribov operator; Bargmann space; Eigenvalue problem; Reggeon field theory.\\
{\bf MR(2010) Subject Classification :} 47B36, 47B39.\\

$\overline{\quad \quad \quad \quad \quad \quad \quad \quad \quad \quad \quad \quad \quad \quad \quad \quad \quad \quad \quad \quad \quad \quad \quad \quad \quad \quad \quad \quad \quad \quad \quad \quad \quad \quad \quad \quad \quad \quad \quad \quad \quad \quad}$ \\

\begin{center}
{\color{blue}{{\Large \bf 1 Introduction}}}\\
\end{center}
\quad\\
Usually, quantum Hamiltonians are constructed as selfadjoint operators; for certain situations, however, non-selfadjoint Hamiltonians are also of importance. In particular, the reggeon field theory (as invented by V. Gribov {\color{blue}[5]}) for the high energy behavior of soft processes is governed  in zero transverse dimension by the non-selfadjoint operator:\\

$\displaystyle{H_{\lambda',\mu,\lambda} = \lambda' A^{*^{2}}A^{2} + \mu A^{*}A + i\lambda A^{*}(A + A^{*})A}$$\hfill { }(1.1)$\\

Where $A$ and $A^{*}$ are the standard Bose annihilation and creation operators that:\\

$[A, A^{*}] = I \hfill { }  (1.2)$\\

and\\

$\lambda'$, $\mu$, $\lambda$ are real numbers ($\lambda'$ is the four coupling of Pomeron, $\mu$ is Pomeron intercept and $\lambda$ is the triple coupling of Pomeron see Gribov {\color{blue}[5]} or the excellent survey of the development of Reggeon theory and its application to hadron interactions at high energies given by Boreskov et al. at 2006 in {\color{blue}[4]}.) and $i^{2} = -1$.\\

We denote the Bargmann space {\color{blue}[3]} by :\\

$\mathbb{B}$ = $\{\phi: I\!\!\!\!C\longrightarrow  I\!\!\!\!C\, entire ; \displaystyle {\int_{I\!\!\!\!C}}\displaystyle{\mid\phi(z)\mid^{2}}e^{-\mid z\mid^{2}}dxdy < \infty \}$ $\hfill { }(1.3)$\\
\vspace{0.5cm}

The scalar product on $\mathbb{B}$ is defined by\\

$<\phi,\psi> = \displaystyle {\int_{I\!\!\!\!C}}\displaystyle{\phi(z)\overline{\psi(z)}e^{-\mid z\mid^{2}}dxdy}\hfill { }(1.4)$\\

and the associated norm is denoted by $\mid\mid . \mid\mid $.\\

$\mathbb{B}$ is closed in $L_{2}(\mathbb{C},d\mu(z))$ where the measure $d\mu(z) = e^{-\mid z\mid^{2}}dxdy$ \\

An orthonormal basis of $\mathbb{B}$ is given by $\displaystyle{e_{n}(z) = \frac{z^{n}}{\sqrt{n!}};  n= 0, 1, . . . .}$\\

In this representation, the annihilation and creation operators are defined by\\

$A\phi(z) = \phi'(z)$ with domain $D(A) = \{\phi \in \mathbb{B}; \phi' \in \mathbb{B}\}$\\

$A^{*}\psi(z) = z\psi(z)$ with domain $D(A^{*}) = \{\psi \in \mathbb{B}; z\psi \in \mathbb{B}\}$\\

Therefore, the Gribov operator $H_{\lambda',\mu,\lambda}$ can be written as\\

$\displaystyle{H_{\lambda',\mu,\lambda}\phi = (\lambda^{'} z^{2} + i\lambda z)\phi''(z) + (i\lambda z^{2} + \mu z)\phi'(z)}$ $\hfill { } (1.4)$\\

where $\phi'$ and $\phi''$ are the first and second derivatives of $\phi(z)$ at $z$.

As the vacum is the trivial eigenvector of $H_{\lambda',\mu,\lambda}$ i.e zero is eigenvalue of $H_{\lambda',\mu,\lambda}$ without interest then the maximal domain of $H_{\lambda',\mu,\lambda}$ is given by:\\

$\displaystyle{D(H_{\lambda',\mu,\lambda}) = \{\phi \in \mathbb{B}_{0}; H_{\lambda',\mu,\lambda}\phi \in \mathbb{B}_{0}\}}$ $\hfill { }(1.5)$\\

where\\

$\displaystyle{\mathbb{B}_{0} = \{\phi \in \mathbb{B};\quad \phi(0) = 0\}}$\\

In Bargmann representation {\color{blue}[3]}, the principal spectral properties of $H_{\lambda',\mu,\lambda}$ have been studied in ref.{\color{blue}[7]} and {\color{blue}[12]}.\\

The mathematical difficulties of this problem come of course from the non-self-adjointness of $H_{\lambda',\mu,\lambda}$. Notice that this non-self-adjointness is regular if $\lambda' \neq 0$ ( the domain of anti-adjoint part is inclued in the domain of adjoint part) and it is a rather wild one if $\lambda' = 0$; the word "wild" meaning here that the domains of the adjoint and anti-adjoint parts are not included in one another, nor is the domain of their commutator.\\

For $\lambda' > 0 ,\mu > 0$, let $\sigma (\lambda',\mu) \neq 0$ be the smallest eigenvalue of $H_{\lambda',\mu,\lambda}$, we show in this paper that $\sigma (\lambda',\mu)$ is positive, increasing and analytic function on the whole real line with respect to $\mu$ and that the spectral radius of $H_{\lambda',\mu,\lambda}^{-1}$ converges to that of $H_{\mu,\lambda, 0}^{-1}$ as $\lambda'$ goes to zero.\\ The above results can be derived from the method used in ({\color{blue} [2]} Commun. Math. Phys. 93, (1984), p:123-139) by Ando-Zerner  to study the smallest eigenvalue $\sigma (0,\mu)$ of $H_{0,\mu,\lambda}$, however as $H_{\lambda',\mu,\lambda}$ is regular perturbation of $H_{0,\mu,\lambda}$ then its study is much more easily. We can exploit the structure of $H_{\lambda',\mu,\lambda}^{-1}$ to deduce the results of Ando-Zerner established on the function $\sigma (0,\mu)$ as $\lambda'$ goes to zero.\\

Now we give an outline of the content of this paper, section by section. In Section 2, we recall some spectral properties of $H_{\lambda',\mu,\lambda}$ and of $H_{0,\mu,\lambda}$ established in {\color{blue}[7]} which are necessary for further development and we recall the results of Ando-Zerner {\color{blue}[2]} on analyticity of $\sigma(0, \mu)$ with respect to the parameter $\mu$.\\
In Section 3, we explicit the inverse of $H_{\lambda',\mu,\lambda}$ on $[0, -i\rho']$ which is given by:\\

$\displaystyle{K_{\lambda',\mu,\lambda}\psi(-iy) = \int_{0}^{\rho'}N_{\lambda',\mu,\lambda}(y, y_{1})\psi(-iy_{1})dy_{1}}$ $\hfill { }(1.6)$\\

where \\

$\displaystyle{N_{\lambda',\mu,\lambda}(y, y_{1}) = \frac{1}{\lambda'y_{1}} e^{\rho'y_{1}}(\rho' - y_{1})^{\delta}\int_{0}^{min(y,y_{1})}e^{-\rho's}( \rho' - s)^{-(\delta +1)}ds}$ $\hfill { }(1.7)$\\

 \n with\\
 .\quad \quad $\rho' = \frac{\lambda}{\lambda'}$, \quad $y \in [0, \rho']$,\quad $s \in [0, y]$, \quad $y_{1} \in [s, \rho']$ and $\psi \in \mathbb{B}_{1}$\\

 where\\

 $\mathbb{B}_{1}$ is the family of all restrictions $\phi_{\mid_{[0, -i\rho']}}$ to $[0, -i\rho']$ of $\phi$ in $\mathbb{B}_{0}$.\\

 This restriction permit us to show in section 4 that $K_{\lambda',\mu, \lambda}$ can extended to Hilbert-Schmidt operator of  $L_{2}([0, \rho'], r(y)dy)$  to  $L_{2}([0, \rho'], r(y)dy)$.\\

 with\\

 $\displaystyle {r(y) = e^{2\rho'y}(1 - \frac{y}{\rho'})^{2\delta}}; \quad \delta = \rho'(\rho + \rho') - 1$.$\hfill { }(1.8)$\\

In section 5, By using the properties of kernel of $K_{\lambda',\mu,\lambda}$ and by applying the Jentzsch theorem {\color{blue}[14]}, we show that the smallest eigenvalue $H_{\lambda',\mu,\lambda}$ is positive, increasing and analytic function on the whole real line with respect to $\mu$ and also that the spectral radius of $H_{\lambda',\mu,\lambda}^{-1}$ converges to that of $H_{0, \mu,\lambda}^{-1}$ as $\lambda'$ goes to zero.\\

\begin{center}

{\color{blue}{{\Large \bf 2 Revisited some spectral properties of $H_{\lambda',\mu,\lambda}$ and $H_{0,\mu,\lambda}$}}}

\end{center}

Now, we begin by reviewing some spectral properties of the operators $H_{\lambda',\mu,\lambda}$ and $H_{0,\mu,\lambda}$ established in {\color{blue} [7]}:\\

(a) For $\lambda' \neq 0,  H_{\lambda',\mu,\lambda}$ has compact resolvent and its  domain $D(H_{\lambda',\mu,\lambda}) = D(S)$\\ where\\
$S = A^{*{2}}A^{2}$ and $D(S) = \{\phi \in \mathbb{B}; A^{*{2}}A^{2}\phi \in \mathbb{B}\}$.\\

(b) For $\lambda' > 0$ and $\mu > 0$, $H_{\lambda',\mu,\lambda}$ and $H_{0,\mu,\lambda}$ are invertible on $\mathbb{B}_{0} = \{\phi \in \mathbb{B};\phi(0) = 0\}$  and $H_{\lambda',\mu,\lambda}^{-1}$ converges to $H_{\mu,\lambda}^{-1}$ as $\lambda'$ goes to  z\'ero.\\

(c) Pour $\lambda' > 0$ and $\lambda' \leq \mu\lambda' + \lambda^{2}$, the spectrum of $H_{\lambda',\mu,\lambda}$  is real and the system of its root vectors of is dense in Bargmann space $\mathbb{B}$\\

(d) For $\mu > 0$, the spectrum of $H_{0,\mu,\lambda}$ is real.\\

(e) For $\mu > 0$, there exists an eigenvalue $\sigma(0,\mu) \neq 0$ in the spectrum of $H_{0, \mu,\lambda}$.\\

We can see {\color{blue}[6]}, {\color{blue}[8]}, {\color{blue}[9]}, {\color{blue}[10]}, {\color{blue}[11]}, {\color{blue} [12]} and {\color{blue}[13]} for another spectral properties of  $H_{\lambda',\mu,\lambda}$ and $H_{0,\mu,\lambda}$.\\

In {\color{blue}[2]}, Ando and Zerner  have  given an explicit form of inverse of $H_{\mu,\lambda}$ on negative imaginary axis; $z = -iy $ with $y > 0$.\\

 In this case , $H_{0,\mu,\lambda}^{-1}$ can be written as integral operator in following form :\\

$\displaystyle{H_{0,\mu,\lambda}^{-1}\psi(-iy) = \int_{0}^{+\infty}N_{0,\mu,\lambda}(y, s)\psi(-is)ds}$ $\hfill { }(2.1)$\\

\n where\\

$\displaystyle{N_{0,\mu,\lambda}(y, s) = \frac{1}{\lambda s}e^{\frac{-s^{2}}{2} - \frac{\mu}{\lambda}s}\int_{0}^{min(y, s)}e^{\frac{u^{2}}{2} + \frac{\mu}{\lambda}u}du}$.$\hfill { }(2.2)$\\

Let $\displaystyle{L_{2}([0, +\infty[, e^{-x^{2} -2\omega x}dx)}$ be the space of square
integrable  functions with respect to the measure $e^{-x^{2} -2\omega x}dx$  then  we have the following result due to Ando-Zerner {\color{blue}[2]}:\\

{\bf Proposition} (Ando-Zerner {\color{blue}[2]})\\

i) $\forall \quad \mu \in \mathbb{C}; \mathcal{R}e \mu \geq \omega$  then $H_{0,\mu,\lambda}^{-1}$ can be extended to Hilbert-Schmidt \\operator of $\displaystyle{L_{2}([0, +\infty[, e^{-x^{2} -2\omega x}dx)}$ to $\displaystyle{L_{2}([0, +\infty[, e^{-x^{2} -2\omega x}dx)}$ .\\

ii) For $\mu > 0$, let $\sigma(0,\mu)$ be the smallest eigenvalue of $H_{0,\mu,\lambda}$ then $\sigma (0,\mu)$\\ is positive, increasing and analytic function on the whole real line with respect to $\mu$.\\

\begin{center}

{\color{blue}{{\Large \bf 3 Explicit inversion of $H_{\lambda',\mu,\lambda}$ on $[0, -i\frac{\lambda}{\lambda'}]$}}}

\end{center}

In all this section we put $\rho' = \frac{\lambda}{\lambda'}$, $\rho = \frac{\mu}{\lambda}$ and $\delta = \rho'(\rho + \rho') -1$ for $\lambda' \neq 0$ and $\lambda \neq 0$.\\

In analogy to Ando-Zener method {\color{blue}[2]}, we begin by to explicit the inverse of  $H_{\lambda',\mu,\lambda}$ on $[0, -i\frac{\lambda}{\lambda'}]$ \\

 Let $\psi \in \mathbb{B}_{0}$ and $\phi \in D(H_{\lambda',\mu,\lambda}$, we consider the equation  $H_{\lambda',\mu,\lambda}\phi = \psi$ witch can be written under following form :\\

$\displaystyle{(\lambda'z^{2} + i\lambda z)\phi''(z) = (i\lambda z^{2}  + \mu z)\phi'(z) = \psi(z)}$ $\hfill { }(3.1)$\\

 Let $\psi \in \mathbb{B}_{0}$ and choosing the straight line connecting $-i\rho' , z \in \mathbb{C}$ parametrized  by \\

 $\displaystyle{\gamma : [0, 1] \rightarrow \mathbb{C}, \gamma(t) = -i\rho' + t(z + i\rho') \quad \quad (\gamma(0) = -i\rho' , \gamma(1) = z)}$\\

 If we define $\displaystyle{\int_{-i\rho'}^{z}\psi(\xi)d\xi:=\int_{\gamma}\psi(\xi)d\xi:= \int_{0}^{1}}\psi(\gamma(t))\gamma'(t)dt$ then the equation (3.1) can be transformed to following integral equation :\\

$\displaystyle{\phi(z) = \frac{1}{\lambda'}\int_{-i\rho'}^{z}e^{-i\rho'\eta}(\eta + i\rho')^{-(\delta +1)} [\int_{-i\rho'}^{\eta}e^{i\rho'\xi}(\xi + i\rho')^{\delta}\frac{\psi(\xi)}{\xi}d\xi]d\eta}$ $\hfill { }(3.2)$\\

However, the integral representation of $\phi(z)$ in this last equation (3.2) is hard to study in $\mathbb{C}$ for some existing results on eigenvalues and eigenfunctions of our operators. To overcome this difficulty, the study of (3.2) is restricted on negative imaginary axis by setting :\\

$ z = -iy$, $u(y) = \phi(-iy)$ and $ f(y) = \psi(-iy)$ with $y \in [0, \rho']$ then (3.1) can be written in the following form:\\

$\displaystyle{(\lambda'y^{2} - \lambda y)u''(y) + (\lambda y^{2} + \mu y) u'(y) = f(y)}$ $\hfill { }(3.3)$\\

As $\psi \in \mathbb{B}_{0}$ then $f(0) = 0$ and if we put $\eta = -is$ and $\xi = y_{1}$ with , $s \in [0, y]$ and $y_{1} \in [s, \rho']$ then (3.3) can be transformed to  following integral equation\\

$\displaystyle{\phi(-iy) = \frac{1}{\lambda'}\int_{0}^{y}e^{-\rho's}(\rho' - s)^{-(\delta +1)}\int_{s}^{\rho'}e^{\rho'y_{1}}(\rho' - y_{1})^{\delta}\frac{\psi(-iy_{1})}{y_{1}}dy_{1}ds}$ $\hfill { }(3.4)$\\

then we get\\

$\displaystyle{K_{\lambda',\mu,\lambda}\psi(-iy) = \int_{0}^{\rho'}N_{\lambda',\mu,\lambda}(y, y_{1})\psi(-iy_{1})dy_{1}}$ $\hfill { }(3.5)$\\

with \\

$\displaystyle{N_{\lambda',\mu,\lambda}(y, y_{1}) = \frac{1}{\lambda'y_{1}} e^{\rho'y_{1}}(\rho' - y_{1})^{\delta}\int_{0}^{min(y,y_{1})}e^{-\rho's}( \rho' - s)^{-(\delta +1)}ds}$ $\hfill { }(3.6)$\\

or\\

$\displaystyle{N_{\lambda',\mu,\lambda}(y, y_{1}) = \frac{1}{\lambda y_{1}} e^{\rho'y_{1}}(1 - \frac{y_{1}}{\rho'})^{\delta}\int_{0}^{min(y,y_{1})}e^{-\rho's}(1 - \frac{s}{\rho'})^{-(\delta +1)}ds}$ $\hfill { }(3.7)$\\

Let $ \Upsilon = min(y,y_{1})$ and $ \displaystyle{\Theta(\Upsilon) = \int_{0}^{\Upsilon}e^{-\rho's}(1 - \frac{s}{\rho'})^{-(\delta +1)}ds}$ \\

then we get the following elementary properties on the function  $\Theta(\Upsilon)$  near to points zero and $\rho'$ respectively:\\

{\bf Lemma 3.1}\\

 Let $ \Upsilon = min(y,y_{1})$ et $ \displaystyle{\Theta(\Upsilon) = \int_{0}^{\Upsilon}e^{-\rho's}(1 - \frac{s}{\rho'})^{-(\delta +1)}ds}$ then

i) $\displaystyle{\Theta(\Upsilon) \equiv \Upsilon}$ near zero.\\

ii)$\displaystyle{\Theta(\Upsilon) \equiv (\rho' -\Upsilon)^{-\delta}}$ near $\rho'$.\\

\quad\\

\begin{center}
{\color{blue}{{\Large \bf 4 Extension of $K_{\lambda',\mu,\lambda}$ to $L_{2}([0, \rho'], r(y)dy)$}}}\\
\end{center}
\quad\\

Let $L_{2}([0, \rho'], r(y)dy)$ be  the space of square
integrable  functions with respect to the weight  $r(y) = e^{2\rho'y}(1 - \frac{y}{\rho'})^{2\delta}; \rho' = \frac{\lambda}{\lambda'}$ and $\delta = \rho'(\rho + \rho') - 1$.\\

By using the lemma 3.1 we obtain the following theorem:\\

{\bf Theorem 4.1}\\

 For $\lambda' > 0$, $\mu > 0$ and $\delta \geq 0$  $K_{\lambda',\mu,\lambda}$ can be extended to Hilbert-Schmidt operator of\\

 $L_{2}([0, \rho'], r(y)dy)$ to $L_{2}([0, \rho'], r(y)dy)$.\\

{\bf Proof}\\

 Let $\displaystyle{r(y) = e^{2\rho'y}(1 - \frac{y}{\rho'})^{2\delta}; \delta = \rho'(\rho + \rho') - 1}$ then \\

$\displaystyle{K_{\lambda',\mu,\lambda}\psi(-iy) = \int_{0}^{\rho'}N_{\lambda',\mu,\lambda}(y, y_{1})\psi(-iy_{1})dy_{1}}$\\

with \\

$\displaystyle{N_{\lambda',\mu,\lambda}(y, y_{1}) = \frac{1}{\lambda y_{1}} e^{\rho'y_{1}}(1 - \frac{y_{1}}{\rho'})^{\delta}\int_{0}^{min(y,y_{1})}e^{-\rho's}(1 - \frac{s}{\rho'})^{-(\delta +1)}ds}$ \\

can be written under the following form :\\

$\displaystyle{K_{\lambda',\mu,\lambda}\psi(-iy) = \int_{0}^{\rho'}\tilde{N}_{\lambda',\mu,\lambda}(y, y_{1})\psi(-iy_{1})r(y_{1})dy_{1}}$ $\hfill { }(4.1)$\\

with \\

$\displaystyle{\tilde{N}_{\lambda',\mu,\lambda}(y, y_{1}) = \frac{1}{\lambda y_{1}\sqrt{r(y_{1})}} \int_{0}^{min(y,y_{1})}e^{-\rho's}(1 - \frac{s}{\rho'})^{-(\delta +1)}ds}$ $\hfill { }(4.2)$\\

or\\

$\displaystyle{\tilde{N}_{\lambda',\mu,\lambda}(y, y_{1}) = \frac{1}{\lambda y_{1}\sqrt{r(y_{1})}} \Theta(min(y,y_{1}))}$ $\hfill { }(4.3)$\\

Now we consider the following integral: \\

$\displaystyle{\mathbb{I} = \int_{0}^{\rho'}\int_{0}^{\rho'}\tilde{N}_{\lambda',\mu,\lambda}^{2}(y, y_{1})r(y)r(y_{1})dydy_{1}}$ $\hfill { }(4.4)$\\

or \\

$\displaystyle{\mathbb{I} = \int_{0}^{\rho'}\int_{0}^{\rho'}\frac{1}{\lambda^{2} y_{1}^{2}} \Theta^{2}(min(y,y_{1}))r(y)dydy_{1}}$ $\hfill { }(4.5)$\\

By applying Fubini theoren then (4.5) can be written as :\\

$\displaystyle{\mathbb{I} = \int_{0}^{\rho'}\int_{0}^{y_{1}}\frac{1}{\lambda^{2} y_{1}^{2}} \Theta^{2}(min(y,y_{1}))r(y)dydy_{1} + \int_{0}^{\rho'}\int_{y_{1}}^{\rho'}\frac{1}{\lambda^{2} y_{1}^{2}} \Theta^{2}(min(y,y_{1}))r(y)dydy_{1}}$ \\

and we have:\\

$\displaystyle{\mathbb{I} = \int_{0}^{\rho'}\int_{0}^{y_{1}}\frac{1}{\lambda^{2} y_{1}^{2}} \Theta^{2}(y)r(y)dydy_{1} + \int_{0}^{\rho'}\int_{y_{1}}^{\rho'}\frac{1}{\lambda^{2} y_{1}^{2}} \Theta^{2}(y_{1})r(y)dydy_{1} = \mathbb{I}_{1} + \mathbb{I}_{2}}$ \\

with\\

$\displaystyle{\mathbb{I}_{1} = \int_{0}^{\rho'}\frac{1}{\lambda^{2} y_{1}^{2}}\int_{0}^{y_{1}} \Theta^{2}(y)r(y)dydy_{1}}$  $\hfill { }(4.6)$\\

and\\

$\displaystyle{\mathbb{I}_{2} = \int_{0}^{\rho'}\frac{\Theta^{2}(y_{1})}{\lambda^{2} y_{1}^{2}}\int_{y_{1}}^{\rho'}r(y)dydy_{1}}$  $\hfill { }(4.7)$\\

Now we remark that \\

i) For $y_{1}$ near zero we have $\displaystyle{\int_{0}^{y_{1}} \Theta^{2}(y)r(y)dy \equiv y_{1}^{3}}$ \\

and\\

ii) For $y_{1}$ near $\rho'$, the function $ y_{1} \rightarrow \displaystyle{\int_{0}^{y_{1}} \Theta^{2}(y)r(y)dy}$ is continuous.\\

then from i) et ii) we deduce that the first integral (4.6) converges. Similarly we verify that the second integral (4.7) converges and consequently the operator $K_{\lambda',\mu,\lambda}$ is Hilbert-Schmidt on $L_{2}([0, \rho'], r(y)dy)$.\\

\begin{center}

 {\color{blue}\Large\bf 5. Application of an generalized version of Jentzsch theorem to $K_{\lambda',\mu,\lambda}$}\\

 \end{center}

In this section we begin by recalling some generalizations of Jentzsch theorem {\color{blue}[14]} which are necessary to prove the main results on the smallest eigenvalue of our operator $H_{\lambda',\mu,\lambda}$.\\

The next theorem is due to Zerner which simplifies considerably the prove of an theorem given in the page 337 of the book Schafer {\color{blue}[17]} which is presented as ``a general version of a classical theorem on kernel operators (theorem of Jentzsch {\color{blue}[14]})".\\

{\bf Theorem 5.1 } ({\color{blue}[17]}, {\color{blue}[18]})\\

 Let $(\mathbb{X},\Sigma, \tau) $ be a measure space with positive measure $\tau$ and $\mathbb{L}_{p}(\tau)$ be the set of all measurable functions on $\mathbb{X}$  whose absolute value raised to the p-th power has finite integral.\\

Let $\mathbb{T}$ be an integral bounded operator defined on $\mathbb{L}_{p}(\tau)$ by a kernel $ N \geq 0$.\\

We suppose that:\\

(i) There exists $n \in \mathbb{N}$ such that $\mathbb{T}^{n}$ is compact operator.\\

(ii) For $\mathbb{S}\in  \sum; \tau(\mathbb{S}) > 0$ and $\tau(\mathbb{X} - \mathbb{S}) > 0$ we have :\\

$\displaystyle{\int_{\mathbb{X}-\mathbb{S}}\int_{\mathbb{S}}N(s,t)\tau(s)d\tau(t) > 0}$ $\hfill { } (5.1)$\\

Then \\

the spectral radius $r(\mathbb{T})$ of integral operator $\mathbb{T}$ is simple eigenvalue  associated to an eigenfunction $f$ satisfying $ f(s) > 0$ $\tau$-almost everywhere.\\

 And if $N(s, t) > 0 \quad\quad \tau\otimes\tau$-almost everywhere, then every eigenvalue $\alpha$ of $\mathbb{T}$ satisfies  $\mid \alpha \mid < r(\mathbb{T})$.\\

Now to apply the above theorem at $K_{\lambda',\mu,\lambda}$, we need of some results given in {\color{blue}[2]} that we recall under theorem form:\\

{\bf Theorem 5.2 (Ando-Zerner {\color{blue}[2]})}\\

Let $\mathbb{E}$ be a Banach space immersed in  $\mathbb{L}_{p}(\tau)$ with more strong norm.\\

 Under the same assumptions that of the above theorem, we suppose that: \\

(i) The operator $\mathbb{T}$ is compact on $\mathbb{E}$  and  $\mathbb{E}$ is invariant by $\mathbb{T}$.\\

(ii) There exists a positive function in  $\mathbb{E}$.\\

Then\\

the spectral radius of $\mathbb{T}$ on $\mathbb{E}$ coincides with its spectral radius on $\mathbb{L}_{p}(\tau)$.\\

Now for the operator (4.1) acting on $L_{2}([0, \rho'], r(y)dy)$ recalled below\\

$\displaystyle{K_{\lambda',\mu,\lambda}\psi(-iy) = \int_{0}^{\rho'}\tilde{N}_{\lambda',\mu,\lambda}(y, y_{1})\psi(-iy_{1})r(y_{1})dy_{1}}$ \\

with \\

$\displaystyle{\tilde{N}_{\lambda',\mu,\lambda}(y, y_{1}) = \frac{1}{\lambda y_{1}\sqrt{r(y_{1})}} \int_{0}^{min(y,y_{1})}e^{-\rho's}(1 - \frac{s}{\rho'})^{-(\delta +1)}ds}$ \\

or\\

$\displaystyle{\tilde{N}_{\lambda',\mu,\lambda}(y, y_{1}) = \frac{1}{\lambda y_{1}\sqrt{r(y_{1})}} \Theta(min(y,y_{1}))}$ \\

 we have the following spectral properties:\\

(i) $K_{\lambda',\mu,\lambda}$ is Hilbert-Schmidt operator on $L_{2}([0, \rho'], r(y)dy)$. In particular, it is a compact operator on $\mathbb{B}_{1}$.\\

(ii) $\mathbb{B}_{1}$ is invariant by $K_{\lambda',\mu,\lambda}$.\\

(ii) Pour $\lambda > 0$, $\tilde{N}_{\lambda',\mu,\lambda}$ is positive.\\

(iii) The function $\displaystyle{\phi(z) = \frac{sin^{2}iz}{iz}}$ is positive and belongs to $\mathbb{B}_{1}$.\\

By applying the above theorems 5.1 and 5.2, we get the following results given under theorem form\\

{\bf Theorem 5.3}\\

(i) For $\lambda > 0$, the operator $K_{\lambda',\mu,\lambda}$ acting on $L_{2}([0, \rho'], r(y)dy)$ has a non-zero eigenvalue.\\

(ii) For $\lambda > 0$, the largest eigenvalue in modulus $\Omega(\lambda',\mu,\lambda)$ of the operator $K_{\lambda',\mu,\lambda}$ is simple, positive and the associated eigenfunction is positive on $\mathbb{R}^{+}$.\\

(iii) $\Omega(\lambda',\mu,\lambda)$  is the largest eigenvalue of $K_{\lambda',\mu,\lambda}$ as an operator acting on $\mathbb{B}_{1}$, where $\mathbb{B}_{1}$ is the family of all restrictions $\phi_{\mid_{[0, -i\rho']}}$ to $[0, -i\rho']$ of $\phi$ in $\mathbb{B}_{0}$.\\

Now we are going to study the convergence of $\displaystyle{r(y) = e^{2\rho'y}(1 - \frac{y}{\rho'})^{2\delta}; \delta = \rho'(\rho + \rho') - 1}$ and the convergence of $K_{\lambda',\mu,\lambda}$ as $\lambda' \rightarrow 0$ i.e as $\rho' \rightarrow +\infty$.\\

If we set :\\

a) $\displaystyle{N_{\rho'}(y, y_{1}) = \tilde{N}_{\lambda',\mu,\lambda}(y, y_{1})\chi_{[0,\rho']\times[0,\rho']}}$ where  $\chi_{[0,\rho']\times[0,\rho']}$ is the indicator function of the interval $[0,\rho']\times[0,\rho']$.\\

b) $\displaystyle{K_{\rho'}\psi(-iy) = \int_{0}^{+\infty}N_{\rho'}(y, y_{1})\psi(-iy_{1})dy_{1}}$\\

c) $\displaystyle{L_{2}([0, +\infty[, r_{\infty}(y)dy)}$ be the space of square
integrable  functions with respect to the measure $\displaystyle{r_{\infty}(y) = e^{-y^{2} -2\rho y}dy}$\\

then we obtain\\

{\bf Lemma 5.4}\\

(i) Let $\displaystyle{r(y) = e^{2\rho'y}(1 - \frac{y}{\rho'})^{2\delta}}$ with $\delta = \rho'(\rho + \rho') - 1$ then\\

$\displaystyle{Lim\quad r(y) = r_{\infty}(y) = e^{-y^{2} - 2\rho y}}$ as  $\rho' \rightarrow +\infty$.\\

(ii) Let $\displaystyle{N_{\rho'}(y, y_{1}) = \frac{1}{\lambda y_{1}\sqrt{r(y_{1})}} \int_{0}^{min(y,y_{1})}e^{-\rho's}(1 - \frac{s}{\rho'})^{-(\delta +1)}ds}$ then\\

$\displaystyle{Lim \quad N_{\rho'}(y, y_{1})}$ = $\displaystyle{N_{\mu,\lambda}(y,  y_{1}) = \frac{1}{\lambda  y_{1}}e^{\frac{- y_{1}^{2}}{2} - \frac{\mu}{\lambda} y_{1}}\int_{0}^{min(y,  y_{1})}e^{\frac{s^{2}}{2} + \frac{\mu}{\lambda}s}ds}$ (the kernel associated to $H_{\mu,\lambda}^{-1}$) as  $\rho' \rightarrow +\infty$.\\

 {\bf Proof}\\

 i) As $\displaystyle{Log\sqrt{r(y)} = \rho'y + \delta Log(1 - \frac{y}{\rho'})}$ then for $\rho' \rightarrow +\infty$ we have:\\

$\displaystyle{Log\sqrt{r(y)}\equiv \rho'y - \delta\frac{y^{2}}{2\rho'^{2}}}$\\

 $\displaystyle{\equiv \rho'y - \rho'(\rho + \rho')\frac{y}{\rho'} + \frac{y}{\rho'} - \rho'(\rho + \rho')\frac{y^{2}}{2\rho'^{2}} + \frac{y^{2}}{2\rho'^{2}}}$\\

$\displaystyle{\equiv \rho'y -\rho y - \rho'y +  \frac{y}{\rho'} - \rho\frac{y^{2}}{2\rho'} - \frac{y^{2}}{2} + \frac{y^{2}}{2\rho'^{2}}}$\\

it follows that:\\

$\displaystyle{lim \quad Log\sqrt{r(y)} = - \frac{y^{2}}{2} - \rho y}$ lorsque  $\rho' \rightarrow +\infty$\\

which establishes the result:\\

$\displaystyle{Lim r(y) = r_{\infty} = e^{- y^{2} - 2\rho y}}$ as  $\rho' \rightarrow +\infty$\\

ii) As in i), similarly if $\displaystyle{A_{\rho'}}$ denotes $\displaystyle{A_{\rho'} = e^{-\rho's}(1 - \frac{s}{\rho'})^{(\delta + 1)}}$, it follows that:\\

$\displaystyle{Lim \quad A_{\rho'} = e^{\frac{s^{2}}{2} + \rho s}}$ as  $\rho' \rightarrow +\infty$\\

which establishes the result:\\

$\displaystyle{Lim \quad N_{\rho'}(y, y_{1})} = \displaystyle{N_{\mu,\lambda}(y,  y_{1}) = \frac{1}{\lambda  y_{1}}e^{\frac{- y_{1}^{2}}{2} - \frac{\mu}{\lambda} y_{1}}\int_{0}^{min(y,  y_{1})}e^{\frac{s^{2}}{2} + \frac{\mu}{\lambda}s}ds}$ the kernel associated to $H_{\mu,\lambda}^{-1}$ as  $\rho' \rightarrow +\infty$.\\

 To deduce the principal result of Ando-Zerner (proposition 1 {\color{blue} [2]}) and the convergence of the spectral radius of $H_{\lambda',\mu,\lambda}^{-1}$ to that of $H_{0,\mu,\lambda}^{-1}$ as $\lambda'$ goes to zero, we need to study $\tilde{N}_{\lambda',\mu,\lambda}$ with respect to the parameters $\lambda'$ and $\mu$ in particular, its analyticity with respect to the parameter $\mu$.\\

{\bf Theorem 5.5}\\

i) Let $\displaystyle{N_{\rho'}(y, y_{1}) = \frac{1}{\lambda y_{1}\sqrt{r(y_{1})}} \int_{0}^{min(y,y_{1})}e^{-\rho's}(1 - \frac{s}{\rho'})^{-(\delta +1)}ds}$ then\\

the function $\rho' \rightarrow N_{\rho'}$ is decreasing.\\

ii) Let $\displaystyle{N_{\rho'}(y, y_{1}) = \frac{1}{\lambda  y_{1}}e^{\frac{- y_{1}^{2}}{2} - \frac{\mu}{\lambda} y_{1}}\int_{0}^{min(y,  y_{1})}e^{\frac{s^{2}}{2} + \frac{\mu}{\lambda}s}ds}$ then\\

$\displaystyle{Lim \quad \int_{0}^{\rho'}N_{\rho'}(y, y_{1})\psi(-iy_{1})dy_{1} = \int_{0}^{+\infty}N_{\mu,\lambda}(y, y_{1})\psi(-iy_{1})dy_{1}}$\\

iii) The integral operator $K_{\rho'}$ can be extended to Hilbert-Schmidt operator of $\displaystyle{L_{2}([0, +\infty[, r_{\infty}(y)dy)}$ to $\displaystyle{L_{2}([0, +\infty[, r_{\infty}(y)dy)}$.\\

iv) On $\displaystyle{L_{2}([0, +\infty[, r_{\infty}(y)dy)}$, the integral operator $K_{\rho'}$ converges to  integral operator $K_{0,\mu,\lambda}$ as $\rho'$ goes to infinity with respect to Hilbert-Schmidt norm.\\

v) Let $\Omega(\lambda',\mu,\lambda)$ be the spectral radius of $K_{\lambda',\mu,\lambda}$  of kernel $N_{\lambda',\mu,\lambda}$ and let $\Omega(0,\mu,\lambda)$ be the spectral radius of $K_{0,\mu,\lambda}$ of kernel $N_{0,\mu,\lambda}$ then\\

$Lim \quad \Omega(\lambda',\mu,\lambda) = \Omega(0,\mu,\lambda)$ as $\lambda' \rightarrow 0$ or $\rho' \rightarrow +\infty$.\\

vi) The functions $\mu \rightarrow \tilde{N}_{\lambda',\mu,\lambda}$ and $\mu \rightarrow \Omega(\lambda',\mu,\lambda)$ are creasing with respect to $\mu$.\\

vii) The function $\mu \rightarrow \Omega(\lambda',\mu,\lambda)$ is analytic on the whole real line with respect to $\mu$ and the function $\mu \rightarrow \Omega(0,\mu,\lambda)$ is also analytic on the whole real line with respect to $\mu$.\\

{\bf Proof}\\

i) By observing that the map $\displaystyle{\rho' \rightarrow e^{\frac{a}{\rho'}}}$ is decreasing where $a$ is an positive constant, we deduce that the map $\rho' \rightarrow  N_{\rho'}$ is also decreasing.\\

ii) We obtain this property from the above lemma.\\

iii) If we consider the integral \\

$\displaystyle{\int_{0}^{+\infty}\int_{0}^{+\infty}N_{\rho'}^{2}\frac{r_{\infty}(y)}{r_{\infty}(y_{1})}dydy_{1}}$\\

then we have:\\

$\displaystyle{\int_{0}^{+\infty}\int_{0}^{+\infty}N_{\rho'}^{2}\frac{r_{\infty}(y)}{r_{\infty}(y_{1})}dydy_{1}}$ = $\displaystyle{\int_{0}^{\rho'}\int_{0}^{\rho'}\tilde{N}_{\lambda', \mu, ,\lambda}^{2}e^{-(y^{2} - y_{1}^{2}) - 2\rho(y - y_{1})}dydy_{1}}$\\

$\displaystyle{\leq C_{\rho'}\int_{0}^{\rho'}\int_{0}^{\rho'}\tilde{N}_{\lambda', \mu, ,\lambda}^{2}dydy_{1}}$ where $C_{\rho'}$ is a constant because the function $\displaystyle{e^{-(y^{2} - y_{1}^{2}) - 2\rho(y - y_{1})}}$ is bounded.\\

As $\displaystyle{\int_{0}^{\rho'}\int_{0}^{\rho'}\tilde{N}_{\lambda', \mu, ,\lambda}^{2}dydy_{1} < +\infty} $ then we deduce the property iii) of this theorem.\\

iv) Let $(\mathbb{X},\Sigma, \tau) $ be a measure space with positive measure $\tau$. An version of the classical monotone convergence theorem of Beppo Levi formulated in terms of functions of the
Lebesgue space $L_{2}(X,\Sigma, \tau)$ reads as follows: If $(f_{n }: n \geq 1)$ is an decreasing sequence (that is, $f_{n}(x) \geq f_{n+1}(x)$ for every $n$ and almost every $x$ in $\mathbb{X}$) of nonnegative,square integrable functions on $\mathbb{X}$ such that $\displaystyle{\int_{X}\mid f_{1}(x)\mid^{2}d\tau < +\infty}$\\

Then\\

$f_{n}$ converges to some square integrable function $f$ both almost everywhere
and in $\displaystyle{L_{2}}$-norm as $n \rightarrow+ \infty$.

By applying this theorem, (ii) of lemma 5.4, (i) and (ii) of theorem 5.5 we deduce that:\\

$\displaystyle{lim\int_{0}^{+\infty}\int_{0}^{+\infty}G_{\rho'}^{2}\frac{r_{\infty}(y)}{r_{\infty}(y_{1})}dydy_{1}}$ $= \displaystyle{\int_{0}^{+\infty}\int_{0}^{+\infty}N_{\mu,\lambda}^{2}\frac{r_{\infty}(y)}{r_{\infty}(y_{1})}dydy_{1}}$ as $\rho' \rightarrow+ \infty$\\

and\\

$K_{\rho'}$ converges to $K(\mu, \lambda)$ as $\rho'$ goes to infinity in Hilbert-Schmidt norm on $\displaystyle{L_{2}([0, +\infty[, r_{\infty}(y)dy)}$.\\

v) Let $\Omega(\lambda',\mu,\lambda)$ be the spectral radius of $K_{\lambda',\mu,\lambda}$  of kernel $N_{\lambda',\mu,\lambda}$ and let $\Omega(0,\mu,\lambda)$ be the spectral radius of $K_{0,\mu,\lambda}$ of kernel $N_{0,\mu,\lambda}$ then\\

As $N_{\lambda',\mu,\lambda}$ is positive decreasing sequence of kernels which converges to the positive kernel $N_{0,\mu,\lambda}$ as goes to infinity we deduce that :\\

$\Omega(\mu,\lambda) \leq Lim \quad \Omega(\lambda',\mu,\lambda) $ as $\rho' \rightarrow + \infty$.\\

 Thus it suffices to show that\\

 $\displaystyle{Lim \quad \Omega(\lambda',\mu,\lambda)\leq \Omega(0,\mu,\lambda)}$ as $\rho' \rightarrow + \infty$.\\

 From the above property iv) we deduce that\\

$\forall \quad n \in \mathbb{N}$, $\displaystyle{\mid\mid K_{\lambda',\mu,\lambda}^{n}\mid\mid^{\frac{1}{n}}}$ $\rightarrow$  $\displaystyle{\mid\mid K_{0,\mu, \lambda}^{n}\mid\mid^{\frac{1}{n}}}$ as $\rho' \rightarrow + \infty$.\\

i.e.\\

$\displaystyle{\forall \quad \epsilon > 0, \exists\quad \rho_{1}; \forall \quad \rho' \geq \rho_{1}}$ we have $\displaystyle{\mid\mid K_{0,\mu, \lambda}^{n}\mid\mid^{\frac{1}{n}} - \epsilon \leq \mid\mid K_{\lambda',\mu,\lambda}^{n}\mid\mid^{\frac{1}{n}} \leq \mid\mid K_{0,\mu, \lambda}^{n}\mid\mid^{\frac{1}{n}} + \epsilon}$\\

set $\rho_{1} = \frac{\mu}{\lambda_{1}^{'}}$ then we get\\

$\displaystyle{\mid\mid K_{\lambda_{1}^{'},\mu,\lambda}^{n}\mid\mid^{\frac{1}{n}}\leq \mid\mid K_{0,\mu, \lambda}^{n}\mid\mid^{\frac{1}{n}} + \epsilon }$.\\

In particular, we get\\

$\displaystyle{\Omega (\lambda_{1}^{'},\mu,\lambda)\leq \Omega(0,\mu, \lambda) + \epsilon }$ $\hfill { } (5.2)$\\

Now as \\

$\displaystyle{\Omega (\lambda',\mu, \lambda) = lim \quad \mid\mid K_{\lambda',\mu, \lambda}^{n}\mid\mid^{\frac{1}{n}}}$ as $n \rightarrow +\infty$\\

i.e.\\

$\displaystyle{\forall \quad \epsilon > 0, \exists\quad n_{0}\in \mathbb{N}; \forall \quad n \geq n_{0}}$ we have $\displaystyle{\Omega (\lambda',\mu, \lambda) - \epsilon \leq \mid\mid K_{\lambda',\mu, \lambda}^{n}\mid\mid^{\frac{1}{n}} \leq \Omega_{(\lambda',\mu, \lambda)} + \epsilon}$\\

By using the above right inequality and the decreasing of the map $\rho' \rightarrow N_{\rho'}$ , we deduce that:\\

$\displaystyle{\mid\mid K_{\lambda',\mu, \lambda}^{n}\mid\mid^{\frac{1}{n}} \leq \Omega(\lambda',\mu, \lambda) + \epsilon \leq  \Omega(\lambda_{1}^{'},\mu, \lambda) + \epsilon}$  $\hfill { } (5.3)$\\

and by (5.2) and (5.3) and for all $\epsilon > 0$, we get:\\

$\displaystyle{lim \quad\Omega_{\rho'}}$ $\leq$ $\displaystyle{\Omega(0,\mu, \lambda) + 2\epsilon}$ as $\rho'\rightarrow +\infty $\\

Then \\

$\displaystyle{lim \quad\Omega(\lambda^{'},\mu, \lambda) = \Omega(0,\mu,\lambda)}$ as $\lambda' \rightarrow 0$.\\

vi) Let $\displaystyle{\tilde{N}_{\lambda',\mu,\lambda}(y, y_{1}) = \frac{1}{\lambda y_{1}\sqrt{r(y_{1})}} \int_{0}^{min(y,y_{1})}e^{-\rho's}(1 - \frac{s}{\rho'})^{-(\delta +1)}ds}$ where\\ $\delta = \rho'(\rho + \rho') - 1 = \rho'^{2} + \rho\rho' -1$\\

then  $\displaystyle{\tilde{N}_{\lambda',\mu,\lambda}(y, y_{1})}$ can be written as \\

$\displaystyle{\tilde{N}_{\lambda',\mu,\lambda}(y, y_{1}) = \frac{1}{\lambda y_{1}}e^{\rho'y_{1}}(1 - \frac{y_{1}}{\rho'})^{\rho'^{2} -1} \int_{0}^{min(y,y_{1})}e^{-\rho's}(1 - \frac{s}{\rho'})^{-\rho'^{2}}[\frac{\rho' - y_{1}}{\rho' - s}]^{\rho\rho'} ds}$ $\hfill { } (5.4)$\\

 As $0 \leq \leq y_{1}$, the map $\displaystyle{\mu \rightarrow [\frac{\rho' - y_{1}}{\rho' - s}]^{\rho\rho'}}$ is creasing with respect to $\mu$ and we get vi).\\

vii) Let $\displaystyle{\delta_{\beta} = \rho'(\frac{\beta}{\lambda} + \rho') - 1}$ and $\displaystyle{L_{2,\beta}([0, \rho'], r_{\beta}(y)dy)}$ be the space of square
integrable  functions with respect to the measure $\displaystyle{r_{\beta}(y) = e^{2\rho'y}(1 - \frac{y}{\rho'})^{2\delta_{\beta}}}$\\

To use the results of chapter VII of Kato's book {\color{blue}[15]} on the operators depending of a parameter, we shall consider the map:\\

$\displaystyle{\mu \rightarrow <\phi, K_{\lambda', \mu, \lambda}\psi > = \int_{0}^{\rho'}\phi(-iy)\int_{0}^{\rho'}\tilde{N}_{\lambda', \mu, \lambda}(y,y_{1})\bar{\psi}(-iy_{1})dy_{1}r_{\beta}(y)dy}$ $\hfill { } (5.5)$\\

 with $\displaystyle{\phi \in L_{2,\beta}([0, \rho'], r_{\beta}(y)dy)}$ and $\displaystyle{\psi \in L_{2,\beta}([0, \rho'], r_{\beta}(y)dy)}$\\

We being by showing that the map defined by (5.5) is continuous with respect to $\mu$ :\\

Let $\displaystyle{\tilde{N}_{\lambda', \beta}}$ be the expression of  $\displaystyle{\tilde{N}_{\lambda', \mu, \lambda}}$  where we have replaced  $\delta $ by $\delta_{\beta}$.\\

Now, we set $\displaystyle{f_{\mu}(y, y_{1}) = e^{2\rho'y}(1 - \frac{y}{\rho'})^{2\delta_{\beta}}\tilde{N}_{\lambda', \mu, \lambda}(y, y_{1})\phi(-iy)\bar{\psi}(-iy_{1})}$ then for $\mu \geq \beta$ we have:\\

$\displaystyle{\mid f_{\mu}(y, y_{1})\mid = e^{2\rho'y}(1 - \frac{y}{\rho'})^{2\delta_{\beta}}\tilde{N}_{\lambda', \beta}(y, y_{1})\mid \phi(-iy)\mid \mid\bar{\psi}(-iy_{1})\mid}$ \\

then the continuity of (5.5) with respect to $\mu$ follows by applying the bounded convergence theorem.\\

Now to show that the map defined by (5.5) is analytic with respect to  $\mu$,  it suffices to observe that for all closed curve we have \\

$\displaystyle{\int_{\gamma}\tilde{N}_{\lambda', \mu, \lambda}d\mu = 0}$ $\hfill { } (5.6)$\\

and to apply the following theorem: \\

{\bf Theorem} (Reed-Simon {\color{blue}[16]}, theorem XII.8) \\

Let $\mathbb{E}$ be a complex Banach space and $L(\mathbb{E})$ be the set of bounded operators on $\mathbb{E}$ with respect to the operator norm.\\

 Let $\mathbb{T}$ be an analytic function $(\mu \in \mathbb{C} \rightarrow \mathbb{T}(\mu) \in L(\mathbb{E}))$ and $\sigma_{0}$  be a simple eigenvalue of $T(\mu_{0})$.\\

Then\\

 There exists an analytic function $\sigma$ of $\mu$  for $\mu$ near $\mu_{0}$ such that  $\sigma(\mu)$ is simple eigenvalue of $\mathbb{T}(\mu)$ and $\sigma(\mu_{0}) = \sigma_{0}$.\\

{\bf Conclusion}\\

For $\lambda' \neq 0$, the operator $H_{\lambda', \mu, \lambda}$  has a rich set of spectral properties desired by physicists.\\

For $\lambda' = 0$,  Ando-Zerner results were a major contribution in the spectral study of the operator $H_{0, \mu, \lambda}$, however the density of its eigenvectors in Bargmann space is open question.\\

\begin{center}
{\color{blue}{\Large{\bf References}}}
\end{center}

\n{\color{blue} [1]}  M.T. Aimar, A. Intissar, and  Paoli,J.-M. ``Quelques nouvelles propri\'et\'es de r\'egularit\'e de l'op\'erateur de Gribov", Comm. Math. Phys. 172, 461–466 (1995) (in French) \\

\n{\color{blue} [2]} T. Ando, and M. Zerner,`` Sur une valeur propre d'un op\'erateur", Commun. Math. Phys., 93, (1984) (in French)\\

\n{\color{blue} [3]} V. Bargmann, ``On Hilbert space of analytic functions and associated integral
transform", Part I, Commun. Pure App. Math., 14, 187-214 (1961)\\

\n {\color{blue}[4]} K.G. Boreskov, A.B. Kaidalov  and O. V. Kancheli,`` Strong interactions at high energies in the Reggeon approch", Phys. Atomic Nuclei, 69 (10), 1765-1780 (2006)\\

\n{\color{blue} [5]} V. Gribov,`` A reggeon diagram technique", Soviet Phys. JETP 26, no. 2, 414-423 (1968)\\

\n{\color{blue} [6]}  A. Intissar,  Le Bellac and M. Zerner,`` Properties of the Hamiltonian of Reggeon field theory", Phys. Lett. B 113, 487-489 (1982)  \\

\n{\color{blue} [7]}  A. Intissar,`` Etude spectrale d'une famille d'op\'erateurs non-sym\'etriques intervenant dans la th\'eorie des champs de reggeons", Commun. Math. Phys. 113, 263-297 (1987) (in French) \\

\n{\color{blue} [8]} A. Intissar, `` Quelques nouvelles propri\'et\'es spectrales de l'hamiltonien de la th\'eorie des champs de reggeons", C.R. Acad. Sci. Paris, t. 308, S\'erie. I, 209-214 (1989) (in French) \\

\n{\color{blue} [9]} A.  Intissar,`` Analyse Fonctionnelle et Th\'eorie Spectrale pour les Op\'erateurs Compacts Non Auto-Adjoints", Editions Cepadues, Toulouse, (1997) (in French)\\

\n {\color{blue}[10]} A. Intissar,`` Analyse de Scattering d'un op\'erateur cubique de Heun dans l'espace de Bargmann", Comm. Math. Phys., 199, 243-256 (1998)  (in French)\\

\n{\color{blue} [11]} A.  Intissar, ``Diagonalization of Non-selfadjoint Analytic Semigroups and Application to the Shape Memory Alloys Operator", J. Math. Anal. Appl. 257, 1-20  (2001)\\

\n{\color{blue} [12]} A.  Intissar, ``Approximation of the semigroup generated by the Hamiltonian of Reggeon field theory in Bargmann space", J. Math. Anal. Appl. 305, 669–689  (2005)\\

\n{\color{blue}[13]} A. Intissar, ``Spectral Analysis of Non-self-adjoint Jacobi-Gribov Operator and Asymptotic Analysis of Its Generalized Eigenvectors", Advances in Mathematics (China), Vol.43, No.x, (2014) doi: 10.11845/sxjz.2013117b\\

\n{\color{blue} [14]} P. Jentzsch, ``Uber Integralgleichungen mit positivem Kern", J. Reine Angw. Moth. 141,  235-244 (1912)  \\

\n{\color{blue} [15]}  T. Kato, ``Perturbation theory for linear operators", Berlin, Heidelberg, New York: Springer (1966)\\

\n{\color{blue} [16]} M. Reed and B. Simon,`` Analysis of operators" (Methods of Modern Mathematical Physics IV). New York: Academic Press (1978)\\

\n{\color{blue} [17]} H. H. Schafer, ``Banach lattices and positive operators", Berlin, Heidelberg, New York: Springer  (1974)\\

\n{\color{blue} [18]} M.  Zerner, `` Quelques propri\'et\'es spectrales des op\'erateurs positifs", Journal of Functional Analysis, 72,  381-417 (1987) (in French)\\

\end{document}